\documentclass{iopart}
\usepackage{amsmath,amssymb}
\newcommand{\ket}[1]{| #1\rangle}                       %
\newcommand{\Van}{{Van\'i\ifmmode \check{c}\else \v{c}\fi{}ek} }
\newcommand{\GLE}{\Gamma_{_{\rm LE} }}                      %
\newcommand{\GAFA}{\Gamma_{_{\rm AFA}} }  
\newcommand{\GAFADR}{\GAFA^{^{\rm DR}}}  
\newcommand{\SAFA}{\Big|\,\overline{O(t)}\,\Big|^2}

\newcommand{\ODR}{O_{_{\rm DR}}(t)}
\newcommand{\OnDR}{O_{_{\rm DR}}(n)}
\newcommand{\sldos}{\sigma_{_{\rm LDOS}}}                 
\newcommand{\equa}[1]{equation~(\ref{#1})}       						%

\usepackage{graphicx,graphics,rotating}	
\usepackage{dcolumn}
\usepackage{bm,pstricks}
\usepackage{hyperref}
\usepackage{mathptmx}
\newrgbcolor{darkblue}{0.165 0.1 0.43}
\newrgbcolor{fbkblue}{0.2304 0.3476  0.5937}
\newrgbcolor{lcolor}{0. 0. 0.75}
\newrgbcolor{gcolor2}{.80 .80 1.}
\newrgbcolor{terra}{0.73 0.28 0.22}
\newrgbcolor{qcggreen}{0.71 0.78 0.68}
\newrgbcolor{qcgyellow}{1. 0.85 0.5}
\newrgbcolor{darkgreen}{0.25 0.46 0.22}
\newrgbcolor{lgrey}{0.62 0.62 0.62}
\newrgbcolor{lred}{0.9 0.55 0.55}
\newcommand{\refi}[1]{}
\newcommand{\refii}[1]{}
\newcommand{\nach}[1]{}
\newcommand{\comm}[1]{} 
\begin{document}
\title[The elusive nature of
the Lyapunov regime]{Loschmidt echo in quantum maps: the elusive nature of
the Lyapunov regime.}
\author{Ignacio Garc\'ia-Mata$^{1,2,3}$ and Diego A. Wisniacki$^4$}
\address{%
$^1$ Departamento de F\'isica, Lab. TANDAR --  CNEA, Buenos Aires, Argentina 
}%
\address{$^2$ Instituto de Investigaciones F\'isicas de Mar del Plata (IFIMAR), CONICET--UNMdP,
Funes 3350, B7602AYL
Mar del Plata, Argentina.}
\address{$^3$ Consejo Nacional de Investigaciones Cient\'ificas y Tecnol\'ogicas (CONICET), Argentina}
\address{%
$^4$ {Departamento de F\'isica, FCEyN UBA, and IFIBA, CONICET,  Pabell\'on 1 Ciudad Universitaria, C1428EGA Buenos Aires, Argentina}
}
\ead{i.garcia-mata@conicet.gov.ar}
\date{\today}%
\begin{abstract}
The Loschmidt echo is a measure of the stability and reversibility of quantum
evolution under perturbations of the Hamiltonian. One of the expected
and most relevant characteristics of this quantity for chaotic systems
is an exponential decay with a perturbation independent decay rate given by 
the classical Lyapunov exponent.
However,  a non-uniform decay  -- instead of the  
Lyapunov regime -- has been reported in several systems. 
In this work we find an analytical semiclassical expression for the 
averaged fidelity amplitude that can be related directly to the 
anomalous -- unexpected-- behaviour of the LE.
\end{abstract}
%
\maketitle
\section{Introduction}
	\label{sec:intro}
Sensitivity to perturbations of quantum evolution
is one of the main reasons for irreversibility in low dimensional 
quantum systems. 
The fidelity  \cite{Peres1984} , 
later dubbed Loschmidt echo (LE)  \cite{Jalabert2001}, 
defined as
\begin{equation}
	\label{eq:mt}
M(t)=|\langle\psi_0 | e^{i t H_\Sigma/\hbar}e^{-i 	t H_0/\hbar}|\psi_0 \rangle|^2
\end{equation}
 was proposed  
to measure such a 
sensitivity. 
It is the overlap 
of an initial state $\ket{\psi_0}$   
evolved forward 
up to time $t$ with a Hamiltonian $H_0$, 
with the same state evolved backward in time 
with a perturbed Hamiltonian $H_\Sigma$. The parameter
$\Sigma$ characterizes the strength of the 
perturbation. Equation~(\ref{eq:mt})  can also be interpreted 
as the overlap at time $t$ of the same state evolved forward in time 
with slightly different Hamiltonians. While the first interpretation 
gives the idea of irreversibility, the second is related to the idea of 
sensitivity to perturbations in the Hamiltonian. 
An important fact is that the LE, and other important related quantities like polarization echoes,
were measured 
in several different experimental setups such as NMR \cite{Pastawski2000,Ryan2005} 
acoustic waves \cite{Taddese2009}, wave guides \cite{Bodyfelt2009}, microwave billiards \cite{Schafer2005,Kober2011}, cold atoms \cite{Andersen2006}, 
Bose-Einstein condensates \cite{Ullah2009}, and quantum chaotic maps implemented with atomic spins \cite{Chau2009}.

In recent years different time and perturbation regimes of the 
LE have been studied in detail \cite{Jalabert2001, Jacquod2001,Gorin2006,Jacquod2009}.
{We first summarize the behaviour as a function of time.
For very short times perturbation theory gives Gaussian decay [see e.g. \cite{Cerruti2002,Cerruti2003,Wisniacki2003}].
This short time transient is followed by an asymptotic regime.
The fundamental difference between quantum chaotic and regular systems lies in this regime.  While for the former the decay is exponential,
 the latter exhibits -- on average -- a power-law dependence in time [see reviews \cite{Gorin2006,Jacquod2009}]. 
Finally there is a saturation given by the effective size of the Hilbert space  \cite{Petitjean2005,gutierrez2009}. 
In addition, we point out that for strongly chaotic systems and very small
perturbations, if at  Heisenberg time the decay has not reached the saturation value, then a crossover to a Gaussian decay is observed \cite{Cerruti2002,Cerruti2003}.}

As a function of the perturbation strength -- for chaotic systems -- 
the decay rate of the LE in the exponential
regime presents two different types of dependence.
For small $\Sigma$
 -- Fermi golden rule (FGR) regime -- the decay rate is given by the width of the local density of states (LDOS), $\sigma_{_{\rm LDOS}}$.
 The LDOS is  the distribution of the overlap squared connecting the set of the unperturbed
 eigenfunctions with the perturbed ones
 and its  width is a measure of the action of a perturbation on the system.
 Then there is a crossover to a perturbation independent regime when $\sigma_{_{\rm LDOS}}> \lambda$, with $\lambda$ the largest classical Lyapunov exponent. In this regime -- usually called Lyapunov regime-- the decay rate of the LE is given by $\lambda$. This simple picture is supported by semiclassical results based on statistical arguments and  by  some numerical 
studies \cite{Jalabert2001,Gorin2006,Jacquod2009,Cucc2002,Cucc2004}. 

Recent works \cite{Andersen2006,Silvestrov2003,Wang2004,Pozzo2007,Natalia2009,Casabone2010}  
have nevertheless shown that the accuracy of the above description of the LE -- which aims to provide a 
universal picture -- is limited. 
Deviations from the perturbation 
independence are found usually  in the form of oscillations 
around $\lambda$. Moreover, there are cases where deviations are considerably large rendering the Lyapunov regime non-existing. 

The goal of this paper is to shed some light on the, sometimes very large, 
deviations from the Lyapunov regime.
We show that, after the initial short-time transient, 
the behaviour of the LE (on average) is strongly 
influenced by the squared average fidelity amplitude (AFA). 
In order to do so we use the semiclassical theory known 
as dephasing representation (DR) \cite{vanicek2003,vanicek2004,Vanicek2006},
and a recent result  \cite{Goussev2008}, derived for local perturbations in billiards, to calculate the decay rate of the AFA for uniformly hyperbolic chaotic systems 
-- in particular quantum maps with classical chaotic counterpart --, in the limit of 
fast decaying correlations -- or very large classical 
Lyapunov exponents --  as a function  of $\Sigma$. 

Our extensive numerical results
show that   -- for quantum maps on the torus --  
the AFA  imprints  
signatures of the perturbation to the decay of
the LE not only in the
FGR regime but also for much stronger perturbations.
For a wide range of perturbation strengths,
the semiclassical expression obtained for  the decay rate of the AFA
correctly reproduces the deviations of 
the LE from the -- perturbation independent -- Lyapunov regime.

Aside from the dependence of the decay regimes on time and perturbation strength, we know that 
the type of initial state should be taken into account \cite{Diego2002}. In particular for semiclassical calculations the initial state must be ``classically meaningful'' \cite{Jacquod2009} [like position states or narrow Gaussian states]. We show numerically that for states which are extended in phase space,  
 as expected the Lyapunov regime is difficult to observe and the decay of the LE is dominated 
-- at least initially -- by the decay of the AFA. 

This paper is organized as follows. In Sec.~\ref{sec:semi} we
introduce the average fidelity amplitude.
In Sec.~\ref{sec:dec} we derive a semiclassical analytical expression  the decay rate of the AFA, 
$\GAFA$.
Taking into account that $\GAFA$ is generally assumed to be equal to  $\sigma_{_{\rm LDOS}}$,
in Sec.~\ref{sec:LDOS} we compare  both quantities.
In Sec.~\ref{sec:numres} our results are tested in a paradigmatic system of
quantum chaos,
a perturbed cat map on the torus. We  show numerically that for
general perturbations
$\sigma_{_{\rm LDOS}}$ and $\GAFA$ can be very different (Sec.~\ref{sec:res1}). 
Then,  we study the
decay of the LE using Gaussian initial states for several values
of $\lambda$ and different perturbation types (Sec.~\ref{sec:Gauss}). 
{We show that with increasing $\lambda$ values the contribution of 
$\GAFA$ to the LE becomes more relevant.
When the initial
state is extended in phase space even for small $\lambda$
the initial decay rate of the LE is 
given by $\GAFA$ (Sec.~\ref{sec:ext})}. We summarize and expose our conclusions
in Sec.~\ref{sec:conc}.

\section{Average fidelity amplitude}
\label{sec:semi}
 All the decay regimes studied previously in the literature 
 -- e.g. \cite{Jalabert2001,Jacquod2001,Cucc2002,Cucc2004,Silvestrov2003}, and reviews \cite{Gorin2006,Jacquod2009} --
 for the LE imply some kind of averaging.
Typically either an average over perturbations or over different initial states can be done. Throughout this
work we do the latter. 
We consider averaging over a number  $n_r$ of 
uniformly distributed -- random -- Gaussian (minimum uncertainty) initial states $\{\ket{\psi_j}\}_{j=1}^{n_r}$. 
Then the average
LE is given by
\begin{equation}
\label{eq:Mt_uno}
\overline{M(t)}=\frac{1}{n_r}\sum_{j=1}^{n_r}|O_j(t)|^2.
\end{equation} 
where 
\begin{equation}
	\label{eq:Oj}
O_j(t)=\langle\psi_j | e^{i t H_\Sigma/\hbar}e^{-i t H_0/\hbar}|\psi_j \rangle
\end{equation}
is just the fidelity amplitude (FA) corresponding to state $\ket{\psi_j}$.
In this section we focus on the squared average fidelity amplitude [labeled AFA in Sect. \ref{sec:intro}]
\begin{equation}
  \label{eq:SAFAdef}
\SAFA=\left|\frac{1}{n_r}\sum_{j=1}^{n_r}O_j(t)\right|^2.
\end{equation}
We remark that expanding \equa{eq:SAFAdef}, $M(t)$ can be expressed as  
$n_r \SAFA$ minus twice the real part of 
 the sum of terms of type $O_j(t)O^{*}_i(t)$ with $j\ne i$.
The  average fidelity amplitude is an interesting quantity on its own being a measurable quantity 
in some of the echo experiments \cite{Schafer2005}.
It is known [e.g. in \cite{Cucc2002,Cerruti2003,Gorin2004,Gorin2006,Gutkin2010}
to decay exponentially like  
\begin{equation}
\label{eq:gafa}
\SAFA \sim \exp[-\GAFA t].
\end{equation}
for chaotic systems.
In general, the calculations seem to induce to the conclusion that
$\GAFA$ is proportional to the width of the LDOS $\sldos$. However in the following 
subsections we show that this result only   
holds either when $\Sigma$ 
is small [with respect to $\hbar$, see Sec.~\ref{sec:res1}] or when
the perturbation acts on a small [say localized] portion of the total  phase space.
In addition we find an analytical approximation $\GAFA$  
which will account for the strange  -- strongly perturbation dependent --  behaviour of $\GLE$.  
In the particular case of quantum maps on the torus, a general study of the AFA and 
$\GAFA$  will be presented elsewhere \cite{pulpo2011}.

\subsection{Semiclassical calculation of the decay rate of the average fidelity amplitude}
	\label{sec:dec}
In this section we follow the procedure introduced for billiard systems
in \cite{Goussev2008} 
to obtain an analytical expression for the decay rate of the AFA, 
but we extend it to a phase space setting (Poincar\'e surface of section).
We include a step-by-step demonstration in order for the paper to be as self contained as possible.
One key element of this derivation is the use of
the dephasing representation (DR) which was introduced  in \cite{vanicek2003,vanicek2004}
to calculate semiclassically the AFA -- a very thorough analysis of the DR is given in \cite{Vanicek2006}. The approach is an alternative
to that of \cite{Jalabert2001} and makes the calculations more manageable.
It  takes advantage of the 
 shadowing theorem \cite{Vanicek2006} and the initial value representation \cite{Miller1970}.
The FA in the DR is given by
\begin{equation}
\ODR=\int dq dp W(q,p)e^{-i\Delta S_t(q,p,\Sigma)/\hbar}
\end{equation}
where  the action difference
\begin{equation}
\Delta S_t(q,p,\Sigma)=-\Sigma\int_0^{t}dt' V(q(t'),p(t'),t').
\end{equation}
is the integral of the perturbation along the classical unperturbed orbit, 
and $W(q,p)$ is the Wigner function of the initial state $\ket{\psi}$.
We want to average over a basis set of initial states. 
The Wigner function of the incoherent sum
of \emph{any} basis set is just a constant, 
so the average fidelity amplitude does not depend on the type of initial states.
Therefore we have
\cite{Vanicek2006} 
\begin{equation}
	\label{eq:On}
\overline{\ODR}=\frac{1}{{\cal V}}\int dq dp\ e^{-i\Delta S_t(q,p,\Sigma)/\hbar}
\end{equation}
 (with ${\cal V}$ the volume of phase space).
Although the derivations in this paper can apply to any dynamical system in a Poincar\'e
surface of section, for simplicity the following calculation will be done assuming we have an
abstract map on a phase space of area equal to one {(${\cal V}=1$)}. Since maps are iterated in discrete steps, the time is 
henceforth represented by an integer $n$.

Let us suppose we have a region in phase space of area $\alpha$ where the trajectories that pass through 
$\alpha$ are affected by the perturbation.
The complement, with area $(1-\alpha)$, is unaffected. 
Now let us divide all 
initial points $(q,p)$ into sets $\Omega_{n'}$ corresponding to trajectories 
that have visited the perturbed region exactly $n'$ times 
after $n$ steps.
Then \equa{eq:On} can be written as {\cite{Goussev2008}}
\begin{equation}
\overline{\OnDR}=\sum_{n'=0}^n O_{n'}(n)
\end{equation}   
with
\begin{equation}
O_{n'}(n)=\int_{\Omega_{n'}}dq dp \exp[-i\Delta S_n(q,p,\Sigma)/\hbar].
\end{equation}
Due to the fact that the map is completely chaotic, 
the probability to hit the perturbed region is approximately equal to the area $\alpha$. 
We can thus take the exponential out of the integral and get {\cite{Goussev2008}}
\begin{equation}
	\label{eq:approx}
O_{n'}(n)\approx 
< \exp [-i \Delta S_{n'}(q,p,\Sigma)/\hbar]>_{\Omega_n}
\int_{\Omega_{n'}} dq dp.
\end{equation}
The integral in \equa{eq:approx} is just the fraction of initial conditions that visit the perturbed 
region exactly $n'$ times after $n$ iterations, so
\begin{equation}
\label{eq:binom}
\int_{\Omega_{n'}}  dq dp=
\left(
\begin{array}{c}
n\\
n'
\end{array}
\right)
\alpha^{n'}(1-\alpha)^{n-n'}.
\end{equation}
{In addition
\begin{eqnarray}
\label{eq:sigmanp}
\Delta S_{n'}(q,p,\Sigma)&=& -\Sigma \sum_{n=1}^{n'}V(q_{n},p_{n})\nonumber \\
&=&-\Sigma\left[V(q_{1},p_{1})+\ldots+V(q_{n'},p_{n'})\right]
\end{eqnarray}
where $q_{n}$ and $p_{n}$ are just the $n$-th iteration of the classical map given an initial condition $q$, $p$.
{For strongly chaotic systems, 
$q_n$ and $p_n$ in \equa{eq:sigmanp}
can be  treated as uncorrelated-random variables  \cite{Goussev2008} {-- at least for $n$ small} -- so 
that the average  of \equa{eq:approx} of the product of exponentials can be expressed as the product of 
averages. 
This assumption would be strictly true in the case the map was a perfect random number generator 
\cite{Falcioni2005}
in other words, in the limit $\lambda\to\infty$.
Therefore, if the average is done over a large number of initial states, comparable to the size of the Hilbert space, 
we get}
\begin{equation}
\label{eq:av1}
 < \exp [-i \Delta S_{n'}(q,p,\Sigma)/\hbar]>_{\Omega_n}\approx
 <\exp [-i \Delta S(q,p,\Sigma)/\hbar]>^{n'},
\end{equation}
with $ \Delta S(q,p,\Sigma)$ the action difference after one step.
Using (\ref{eq:approx}), (\ref{eq:binom}), and (\ref{eq:av1}) in 
\equa{eq:On} we obtain the amplitude of the AFA  
\begin{eqnarray}
	\label{eq:Ondc}
\overline{\OnDR}&\approx&\sum_{n'}^{n}
\left(
\begin{array}{c}
n\\
n'
\end{array}
\right)
(\alpha < e^{-i\Delta S(q,p,\Sigma)/\hbar}> )^{n'}
(1-\alpha )^{n-n'}\nonumber\\
&=&
(1-\alpha(1- < e^{-i\Delta S(q,p,\Sigma)/\hbar}>))^{n}.
\end{eqnarray}
This exponential decay can be rewritten as
\begin{equation}
	\label{eq:exp}
\overline{\OnDR}\sim e^{-\Gamma n},
\end{equation}
with 
\begin{equation}
	\label{eq:gammandc}
\Gamma=- \ln (1-\alpha(1-< e^{-i\Delta S(q,p,\Sigma)/\hbar}>)).
\end{equation}
The decay rate of the squared AFA
\begin{equation}
\label{eq:gammafull}
\GAFADR=2\Re \Gamma= 
-2\ln\left|(1-\alpha(1- < e^{-i\Delta S(q,p,\Sigma)/\hbar}>))\right|
 \end{equation}
is then obtained by squaring the modulus of \equa{eq:exp}. 

The previous derivation makes \equa{eq:gammafull} `strictly' valid in the case of local perturbations  and uniformly hyperbolic, chaotic systems or in the case $\lambda \to \infty$. 
For local perturbations ($\alpha\ll 1$) a further 
simplification 
can be made by series expansion in $\alpha$  to obtain an expression equivalent to the one obtained in \cite{Goussev2008,Diego2010}.

We here conjecture that \equa{eq:gammafull} gives a good approximation for the decay rate
 of the AFA -- for  \emph{any} $\alpha$ and $\lambda$, at least for small times -- based mainly on three arguments.  
The first being that in \cite{Diego2010} a 
very similar derivation leading to the width of the local density of states is shown to hold for arbitrary 
values of $\alpha$ [see next section]. The second is that 
we do know that for $t=1$ there is no need for an $\alpha\ll 1$ approximation in 
 \equa{eq:av1} and neither is there need for such approximation in the limit 
 $\lambda\to\infty$. So, \equa{eq:av1} will be valid for
 larger times as $\lambda$ increases. After that there is a crossover to 
 Lyapunov decay [which is observed e.g. in Fig.~\ref{f4figs}, bottom left panel]. The exact time of crossover is difficult to determine because it depends on the form of the perturbation \cite{pulpo2011}.
The last reason  supporting our conjecture is the eloquent 
numerical evidence that we present further in the paper [see section Sec.~\ref{sec:numres}]. 
Particularly, we show that, when the initial states are extended in phase space, e.g. a position state (on the torus), then \emph
{even} for smaller $\lambda$ and $\alpha\to 1$ (global perturbation) the dominating decay rate -- at least for 
initial times -- is $\GAFADR$ obtained in \equa{eq:gammafull}. This last point is illustrated in Sec.~\ref{sec:ext}.
\\

\subsection{{Relation between the local density of states and the fidelity amplitude}}
\label{sec:LDOS}
In the bibliography of the LE \cite{Gorin2006,Jacquod2009} 
it is found that the decay rate for 
small perturbation values -- FGR regime-- originates from the AFA and  
is given by the width of the LDOS ($\sigma_{_{\rm LDOS}}$). Its dependence
with $\Sigma$ is quadratic. 
It is the aim of this section to compare
$\GAFA$ and $\sigma_{_{\rm LDOS}}$ for any perturbation value.

By definition
the LDOS gives the distribution of the perturbed eigenstates in the 
unperturbed eigenbasis. 
It is well known that 
for small $\Sigma$ the shape of the LDOS is a Lorentzian { \cite{Wigner1955}}. 
In \cite{Diego2010} it is shown 
semiclassically
that the shape  
-- in the case of hyperbolic maps and billiards with a deformed boundary as perturbation --
is a Lorentzian   [Breit-Wigner] distribution 
\begin{equation}
L(\omega,\gamma)=\frac{\gamma}{\pi(\omega^2+\gamma^2)},
\end{equation}
for perturbations of arbitrarily high intensity,
with 
\begin{equation}
	\label{eq:gammaldos}
\gamma=\alpha (1-\Re < e^{-i\Delta S(q,p,\Sigma)/\hbar}>).
\end{equation}
The result is obtained
for a perturbation acting on small region of area $\alpha$ and then it is extended to
any value $\alpha$.
The width of the LDOS can be defined as
\begin{equation}
	\label{eq:sigmasc}
\sigma_{_{\rm LDOS}}^{{\rm sc}}=2\,\gamma,
\end{equation}
which corresponds to a distance around the mean value containing approximately 
$70\%$ of the probability 
\footnote{In fact, as seen in \cite{Diego2010}, the correct factor for the distance around the mean
value to be exactly equal to $70\%$ is 1.963. Taking the factor equal to 2 gives a distance around the mean 
containing 
exactly $70.48\%$ of the probability. Since this choice is in a sense arbitrary, for the sake of simplicity, we 
choose the latter.}.

We can now establish the \emph{actual} relationship between the  width of the LDOS and the AFA to the decay
rate of the LE. In the case where the perturbation is local in phase space ($\alpha\ll 1$) from 
Eqs.~(\ref{eq:gammandc}) and (\ref{eq:gammaldos})  we have  {\cite{Goussev2008,Diego2010}}
\begin{eqnarray}
\label{eq:gamma}
\Gamma&\approx& \alpha(1-< e^{-i \Delta S(q,p)/\hbar}>),\\
\label{eq:gamma2}
\GAFADR&=&2\Re \Gamma \equiv 2\gamma =\sigma_{_{\rm LDOS}}^{{\rm sc}}.
\end{eqnarray}
However, 
with arbitrary $\alpha$ the relation of \equa{eq:gamma2} [$2\Re \Gamma \equiv 2\gamma$] will not, in general, be true. 
From \equa{eq:gammafull}  it is evident that 
due to the logarithm, there can appear diverging values. 

\section{Numerical results}
\label{sec:numres}
In this section we take a concrete system 
and perform numerical simulations to 
gauge 
 the analytical results obtained for
$\GAFADR$. First we check its relation with $\sldos$. Next
we do extensive calculations of the decay rate $\GLE$ changing various 
parameters like the Lyapunov exponent and the type of perturbation.
We will also assess the influence of the type of initial states used for the calculations. Notably,
for initial states which are extended in phase space (e.g. squeezed states and position states) we show that
$\GAFADR$ reproduces  with high accuracy the first decay rate of the LE.
We note that after this decay it is possible -- depending on the
values of both $\hbar$ and $\lambda$ -- to 
observe a decay rate given by the Lyapunov exponent.

For the numerical simulations we use quantum maps on the torus.
Quantum maps are very simple systems that represent a useful tool to explore
some of the main properties of classical and quantum chaotic systems \cite{Hannay1980, Balazs1987,Keating1991}.
Furthermore, many quantum maps have been implemented experimentally 
(e.g \cite{Raizen1995, Weinstein2002, Schlunk2003,Henry2006,Chabe2008}).

The quantization of the torus requires imposing periodic boundary conditions leading to a finite dimensional 
Hilbert space of dimension $N$. The associated Planck constant is $\hbar=1/2\pi N$. 
Position basis $\{q_i\}_{i=0}^{N-1}$ (with $q_i\equiv i/n$) and momentum basis $\{p_i\}_{i=0}^{N-1}$ (with $p_i\equiv i/n$) 
are related by the discrete Fourier transform (DFT). 
In such a setting, the evolution operator of 
a quantum map is then a $N\times N$ unitary matrix $U$.
The operator $U$ of the maps we choose to work with can be split into two operators as
\begin{equation}
	\label{eq:Qmap}
U=e^{i 2\pi NT(p)}e^{-i 2\pi NV(q)}
\end{equation}
\begin{figure}[htbp]
\begin{center}
\includegraphics[width=0.9\linewidth]{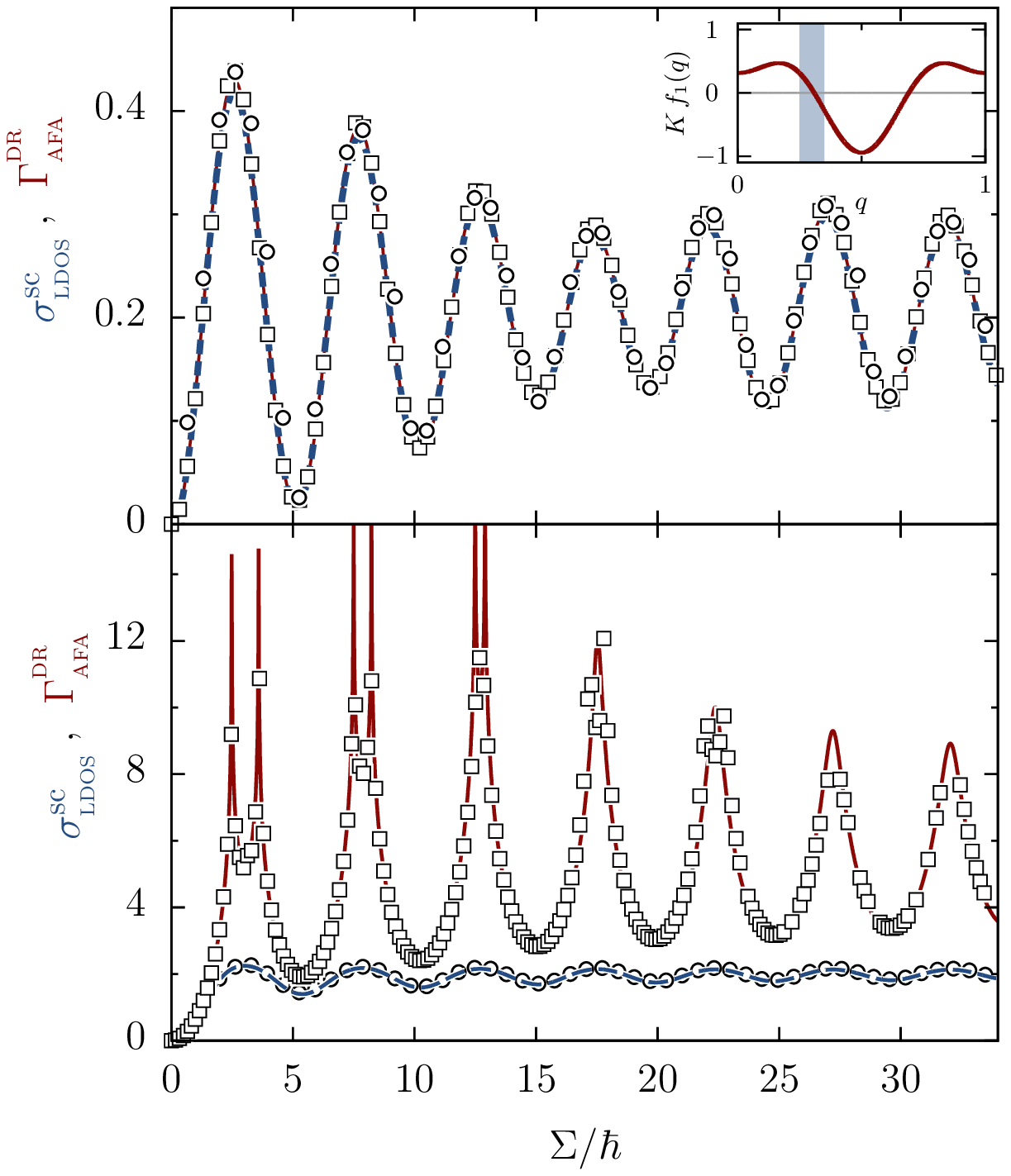} 
\caption{(colour online).
Decay rate of the AFA compared to the width of the LDOS.
In both panels the solid (red) line is $\GAFADR$ from \equa{eq:gammafull} and the dashed (blue) 
line is $\sigma_{_{\rm LDOS}}^{\rm sc}$ of \equa{eq:sigmasc}.
Square symbols represent $\GAFA$ calculated from 
data obtained evolving $n_r=10^4$ initial states for the quantum map corresponding to the classical map of  \equa{eq:pcat} with $a=b=20$, and $f_1(q)$ of \equa{eq:pert}. Hilbert space size is $N=2^{18}$. Circles correspond to the exact calculation of $\sldos$.
 Top panel: the perturbation is localized in the interval $q\in [0.25,0.35]$. Bottom panel: the perturbation acts everywhere. Inset: $K f_1(q)$;  
the shaded region shows where the perturbation was actually applied.
\label{fig-LDOS}
}
\end{center}
\end{figure}
This form is characteristic of periodic delta kicked systems -- like the standard map \cite{Chirikov1979}, the kicked Harper map \cite{Leboeuf1990}, or the sawtooth map. The corresponding classical map can be written as follows
\begin{equation}
\left.
\begin{array}{lcl}
\bar{p}&=&p-\frac{dV(q)}{dq}\\
 \bar{q} &=& q-\frac{dT(\bar{p})}{d\bar{p}}
 \end{array}
\right\} \  ({\rm mod}\ 1).
 \end{equation}
 In particular,
for this contribution we consider the perturbed cat map [see e.g. \cite{Basilio1995}]
 \begin{equation}
 \begin{array}{lcl}
\bar{p}&=&p+a\,q + K f(q) \\
\bar{q}&=&q+b\,\bar{p}+ K' g(\bar{p})\\
\end{array}
\  ({\rm mod}\ 1), \label{eq:pcat}
\end{equation}
with $a$ and $b$ positive integers. 
For $K,\, K'\ll 1$ this map is conservative, uniformly hyperbolic and completely chaotic and  the largest Lyapunov exponent is 
$\lambda\approx\ln((2+ab+\sqrt{ab(4+ab)})/2)$.
The perturbation allows the quantum perturbed cat map to have the generic spectral properties of chaotic systems. The periodicities arising from the arithmetic symmetries
of the map are eliminated for $K,\, K'\ne 0$ \cite{Keating1991,Basilio1995}. In particular the numerical results
shown in this work correspond to three different perturbations: two nonlinear shears in momentum only
\begin{eqnarray}
	\label{eq:pert}
f_1(q)&=& 2\pi(\cos(2\pi q)-\cos(4\pi q))\\
	\label{eq:pert2}
f_2(q)&=& 2\pi \sin(2\pi q)
\end{eqnarray}
with $g(\bar{p})=0$, and the last one composed of $f_1(q)$ above and a nonlinear shear in position 
\begin{equation}
	\label{eq:pert3}
g(\bar{p})=2\pi (\sin(6\pi \bar{p})+\cos(2\pi\bar{p})).
\end{equation}
with $K'=K\,$�in \equa{eq:pcat}.
We consider three different perturbations to illustrate more strongly that the analytical results 
obtained are robust and that $\GAFADR$ correctly reproduces the decay rate of the LE when the 
system is very chaotic.

\subsection{Width of the local density of states vs. decay rate of the fidelity amplitude}
\label{sec:res1}

The results obtained in Sec.~\ref{sec:LDOS}
indicate that indeed there should be a strong correspondence between
$\sigma_{_{\rm LDOS}}$ and $\GAFA$ for perturbations acting on a reduced 
portion of phase space of area $\alpha$ (with $\alpha\ll 1$). In contrast, for
large alpha -- in particular $\alpha\to 1$ -- the two quantities can differ
greatly.

To expose this property, 
in figure~\ref{fig-LDOS} (top) we compare both $\sigma_{_{\rm LDOS}}$ (circles), 
$\sigma_{_{\rm LDOS}}^{{\rm sc}}$ (dashed blue line), with
$\GAFADR$ (solid red line) for 
the perturbation of \equa{eq:pert} restricted to act locally 
on a region $0.25<q<0.35$ [see inset].
As expected  the three curves are very close.
We note that to take into
account the fact that the spectrum of the cat map is periodic because
of a compact phase space, a small correction must be introduced to the
semiclassical approximation $\sigma_{sc}$ of \equa{eq:sigmasc} (see \cite{Diego2010}).

On the other hand, 
we compare {$\sigma^{\rm exact}_{_{\rm LDOS}}$} and  $\sigma_{_{\rm LDOS}}^{{\rm sc}}$ with {$\GAFADR$ (same 
symbols as before, and we include $\GAFA$ -- $\square$ symbols, computed
from $n_r= 10^4$ and $a=b=20$)}
for a perturbation acting
on the whole phase space [figure~\ref{fig-LDOS} -bottom].
In the limiting case $\alpha\to 1$  we have
\begin{equation}
\label{eq:alpha1}
{\GAFADR}=-2\ln \left|< e^{-i\Delta S(q,p)/\hbar}>\right|.
\end{equation} 
This leads to divergences when the argument of the logarithm approaches zero.
In figure~\ref{fig-LDOS} (bottom), we see that
in the global case {$\GAFADR$} has very high (diverging) peaks
while $\sigma_{_{\rm LDOS}}$ oscillates with much smaller -- bounded -- amplitude.
{Only for small perturbations -- $\Sigma/\hbar< 1$ -- both quantities are approximately equal and proportional
to $\Sigma^2$ (FGR).}
We have checked that 
for all the types of perturbation the results agree with the prediction.
%
\begin{figure}[htbp]
\begin{center}
\includegraphics[width=0.9\linewidth]{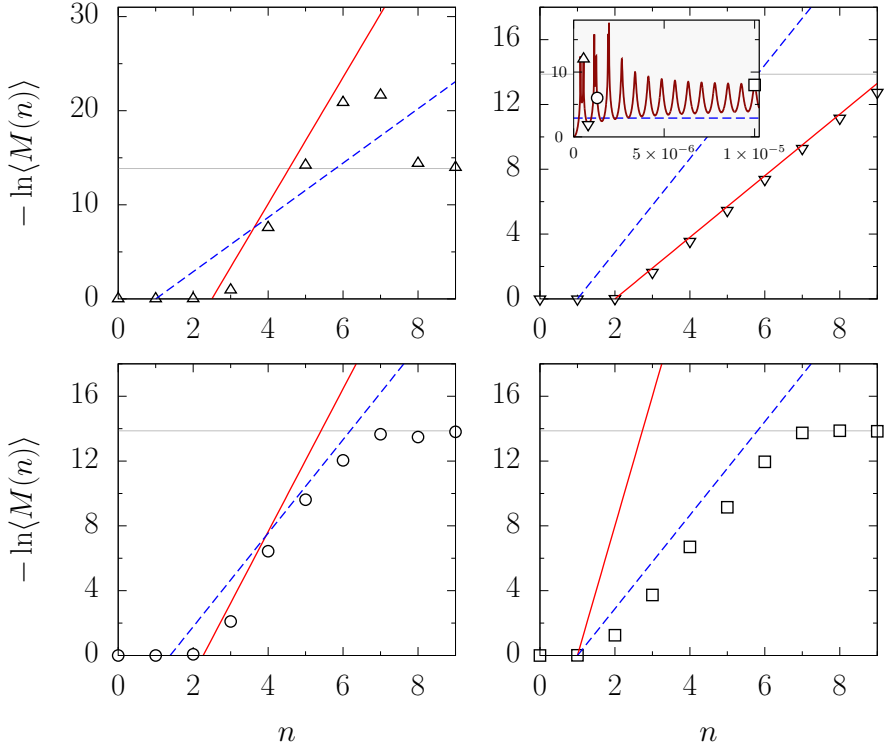} 
\caption{(colour online). LE as a function of time ($n$) for four different values of $\Sigma$ for  $N=2^{20}$, $a=b=4$ ($\lambda\approx2.887$) and using the map (\ref{eq:pcat}) with 
$K f_1(q)$, $K'=0$.
The perturbation values  are:  ($\bigtriangleup$) $\Sigma=4\times 10^{-7}$; 
 ($\bigtriangledown$) $\Sigma=8\times 10^{-7}$; 
 ({\large $\circ$}) $\Sigma=1.35\times 10^{-6}$; 
($\square$) $\Sigma=1\times 10^{-5}$.
The slope of the solid red  is $\GAFADR$ for the corresponding perturbation -- indicated with corresponding symbols in the inset -- and the slope of the dashed blue line 
is $\lambda$. The solid (thin) horizontal grey  line indicates the saturation value ($\ln N$). In the inset
we show $\GAFADR$ calculated from \equa{eq:gammafull} and perturbation $f_1$ of \equa{eq:pert}.
The points indicate the perturbation values taken situated at the corresponding value of $\GAFADR$.
\label{f4figs}
}
\end{center}
\end{figure}
\begin{figure}[htbp]
\begin{center}
\includegraphics[width=0.9\linewidth]{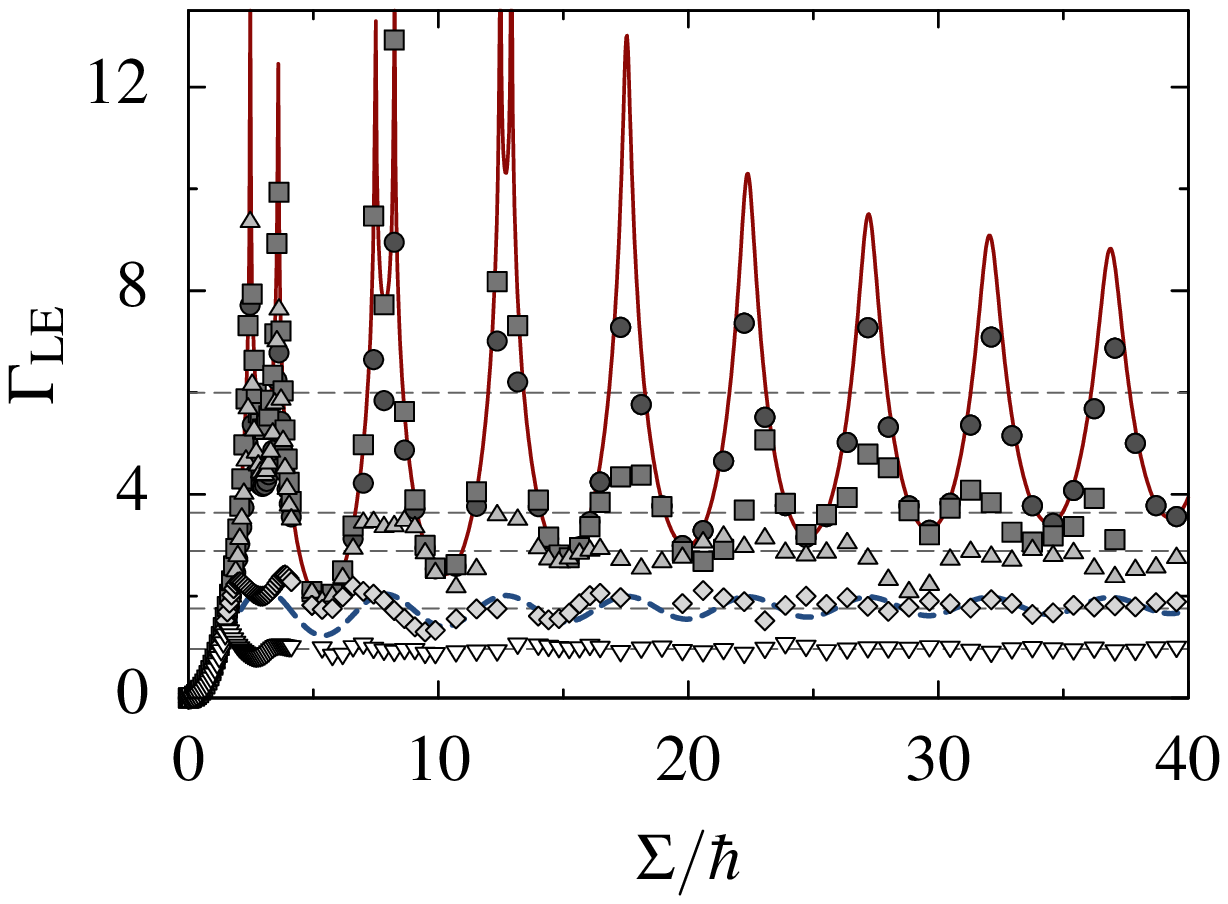} 
\caption{(colour online).
Decay rate $\Gamma_{\rm LE}$ as a function of the rescaled perturbation $\Sigma/\hbar$ for the quantum perturbed cat map with $N=2^{18}$ and for different values of $a$ and $b$ (different values of $\lambda$, indicated by grey the dashed lines). ($\bigtriangledown$) $a=b=1$, $\lambda\approx 0.96$; ({\Large $\diamond$}) $a=b=2$, $\lambda\approx 1.76$; ($\triangle$) $a=b=4$, $\lambda\approx 2.887$; ($\square$) $a=b=6$, $\lambda\approx 3.637$;  ({\large $\circ$}) $a=b=20$, $\lambda\approx 5.996$. The perturbation is $f_1((q)$ [\equa{eq:pert}]
The solid red line corresponds to the $\GAFADR$, the solid blue line is $\sigma_{_{\rm LDOS}}^{{\rm sc}}$
Averages were done over 2048 initial states.
\label{fig:eco}
}
\end{center}
\end{figure}
\\

\subsection{Decay rate of the Loschmidt echo for Gaussian initial states}
\label{sec:Gauss}
For the numerical calculations of the LE we used straightforward quantum 
propagation of the map $U$ -- \equa{eq:Qmap}. The propagation of this map is 
quite efficient
due to the fact that one of the operators is diagonal and the other can be 
implemented using the fast Fourier transform. 
Given an initial state $|\psi_0\rangle$ we computed 
\begin{equation}
M(t)=|\langle \psi_0|(U^\dag_{K_2})^n(U_{K_1})^n|\psi_0\rangle|^2
\end{equation}
where  $K_1,\ K_2$ refer to the perturbation parameter $K$ ($=K'$) in the quantized version of the
map of \equa{eq:pcat} -- we take them 
slightly different-- and $\Sigma$ is defined by
\begin{equation}
\Sigma\equiv |K_2-K_1|.
\end{equation}
As initial state $\psi_0$ we chose coherent -- Gaussian -- states 
\begin{equation}
	\label{eq:coh}
\psi_0(q)=\left({\frac{1}{\pi \hbar \xi^2}}\right)^{1/4}\,\exp\left[\frac{i}{\hbar}p(q-q)-\frac{(q-q)^2}{2 \hbar \xi^2}\right],
\end{equation}
where  $\xi$ is the dispersion in $q$. In this section we chose $\xi=1$ so the state is a symmetric Gaussian in phase space. The periodic boundary conditions that the torus geometry imposes need to be taken into account [left out of \equa{eq:coh} for the sake of simplicity]. 

\begin{figure}[htbp]
\begin{center}
\includegraphics[width=0.9\linewidth]{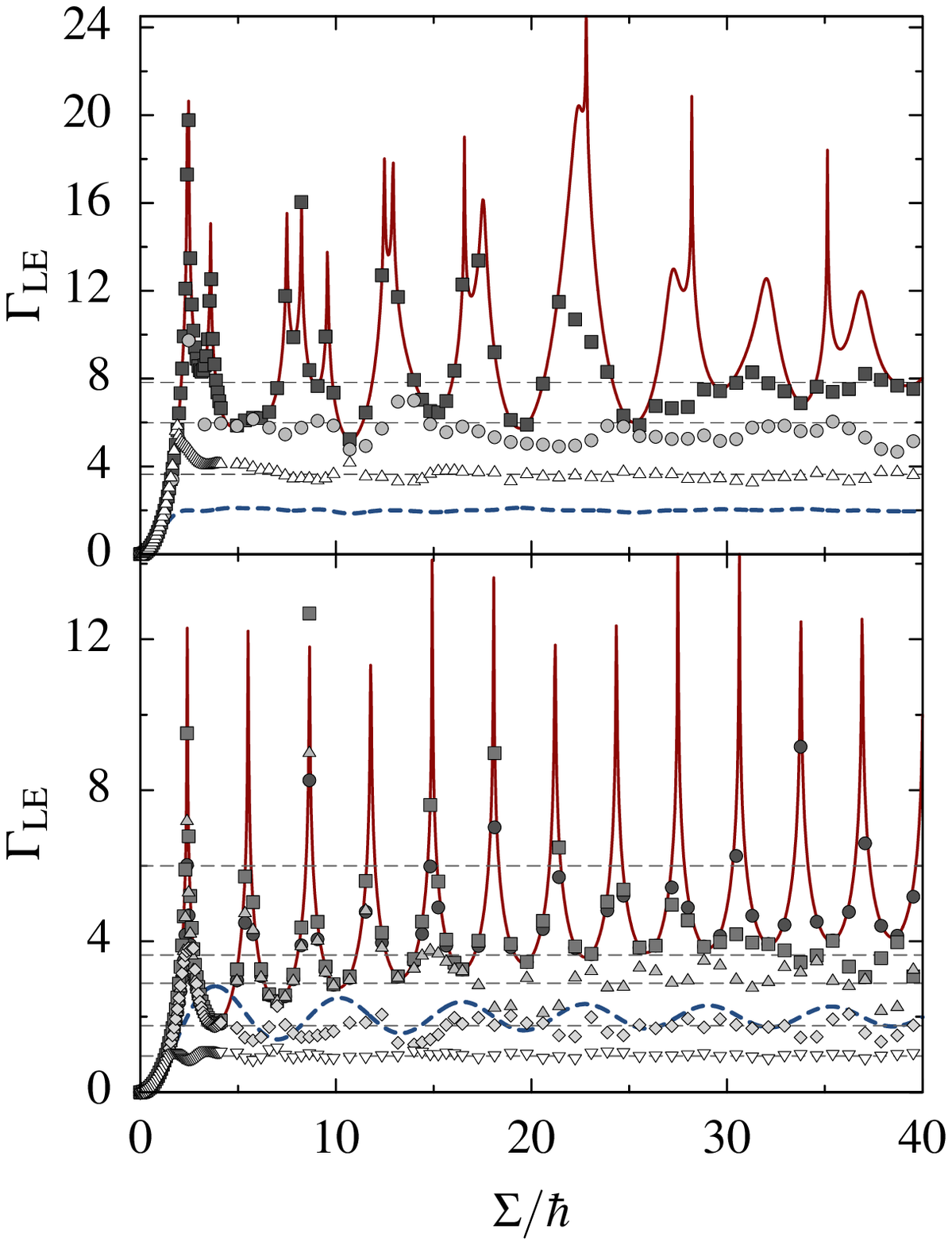} 
\caption{(colour online).
Decay rate $\Gamma_{\rm LE}$ as a function of the rescaled perturbation $\Sigma/\hbar$ 
for the quantum perturbed cat map with $N=2^{18}$, and for two different perturbations 
[see Eqs.~(\ref{eq:pcat}), (\ref{eq:pert}), (\ref{eq:pert2}),(\ref{eq:pert3}) ]: 
{\bf (top)}
$K f_1(q)$, $K' g(\bar{p})$ and {\bf (bottom)} $K f_2(q)$.
  The different symbols represent 
  different values of $a$ and $b$ (different values of $\lambda$, indicated by the dashed lines).
{\bf (top):} ($\bigtriangleup$) $a=b=6$, $\lambda\approx 3.637$;
 ({\large $\circ$}) $a=b=20$, $\lambda\approx 5.996$; 
 ($\square$) $a=b=50$, $\lambda\approx 7.824$.
{\bf (bottom):} ($\bigtriangledown$) $a=b=1$, $\lambda\approx 0.96$; ({\Large $\diamond$}) $a=b=2$, $\lambda\approx 1.76$; ($\triangle$) $a=b=4$, $\lambda\approx 2.887$; ($\square$) $a=b=6$, $\lambda\approx 3.637$;  ({\large $\circ$}) $a=b=20$, $\lambda\approx 5.996$. 
The solid red line corresponds to $\GAFADR$, the dashed blue line is $\sigma_{_{\rm LDOS}}^{{\rm sc}}$
Averages were done over 2048 initial states (Gaussian).
\label{fig:eco2}
}
\end{center}
\end{figure}
We did full quantum simulations with up to dimension $N=2^{20}$.
We needed large dimension in order to have enough points, before saturation, to fit the decay rate
when using maps with large $\lambda$.
In figure~\ref{f4figs} we illustrate
the different alternatives of the decay of the LE as a function of time in the exponential regime.  We show that the decay rate can be  $\GAFADR$ (top-left, top-right)
$\lambda$ (bottom-right),
or a combination of both (bottom-left). 
A time evolution with both decay rates was also observed in \cite{wang2005}
where, contrary to the results found here, 
the authors conclude that the larger decay rate comes from the fluctuations
in the finite-time Lyapunov exponent.
The inset shows $\GAFADR$ as a function of the perturbation, and the points indicate the values of the perturbation for which the LE is plotted.  
For the following figures in the case where two decay rates would be visible, 
we take into account the one corresponding to smaller times {[the crossover behaviour for longer 
times will be dealt with elsewhere \cite{pulpo2011}].}

Next, in figure~\ref{fig:eco} we computed the decay rate $\Gamma_{\rm LE}$,   as
a function of the rescaled perturbation $\Sigma/\hbar$,
for many values of $a,b$ [see \equa{eq:pcat}]
 -- i.e. different values of $\lambda$ and  with perturbation $f_1(q)$.
 We chose  $N=2^{18}$ (we checked some of the results up to $N^{20}$) and averaged over $2048$ initial states.
 For small values of $\lambda$ we obtain the typical result: quadratic dependence for
 small $\Sigma/\hbar$ followed by a constant equal to $\lambda$ (see figure~\ref{fig:eco}, $\bigtriangledown$), 
 the Lyapunov regime.
 As $\lambda$ is increased, the Lyapunov regime becomes less visible and oscillations start to appear.
 Similar oscillations were also observed in  
 \cite{Wang2004,Pozzo2007,Natalia2009,Casabone2010}.
 For large $\lambda$, after the quadratic regime, the oscillations of $\GLE$ follow those of 
 $\GAFADR$ over a finite range. 
 We support this argument with figure~\ref{fig:eco2}, where we show the same as in figure~\ref{fig:eco} for 
 the two other types of perturbations [a combination of (\ref{eq:pert}) with (\ref{eq:pert3}) on the top panel
 and (\ref{eq:pert2}) on the bottom panel]. 
In both cases we see that as $\lambda$ increases, $\GLE$ tends to follow
$\GAFA$ as a function of $\Sigma$. 
 
In both Figs.~\ref{fig:eco} and \ref{fig:eco2}, there seems to be 
three clear regimes: if $\Sigma/\hbar<1$ then we have the FGR regime
where $\GLE=\sldos\propto(\Sigma/\hbar)^2$; this regime is followed by an intermediate regime --which depends on $\lambda$ --  where $\GLE$ is very influenced by $\GAFADR$. Finally for very large $\Sigma/\hbar$ 
the Lyapunov regime appears. 
The crossover between the last two regimes needs
further study, which would include a general understanding of the AFA as a function of the perturbation strength for moderate $\lambda$ values and also understanding the contribution of the ``cross'' terms in the expansion of $\SAFA$ \cite{pulpo2011}. }

\begin{figure} 
\begin{center}
\includegraphics[width=0.9\linewidth]{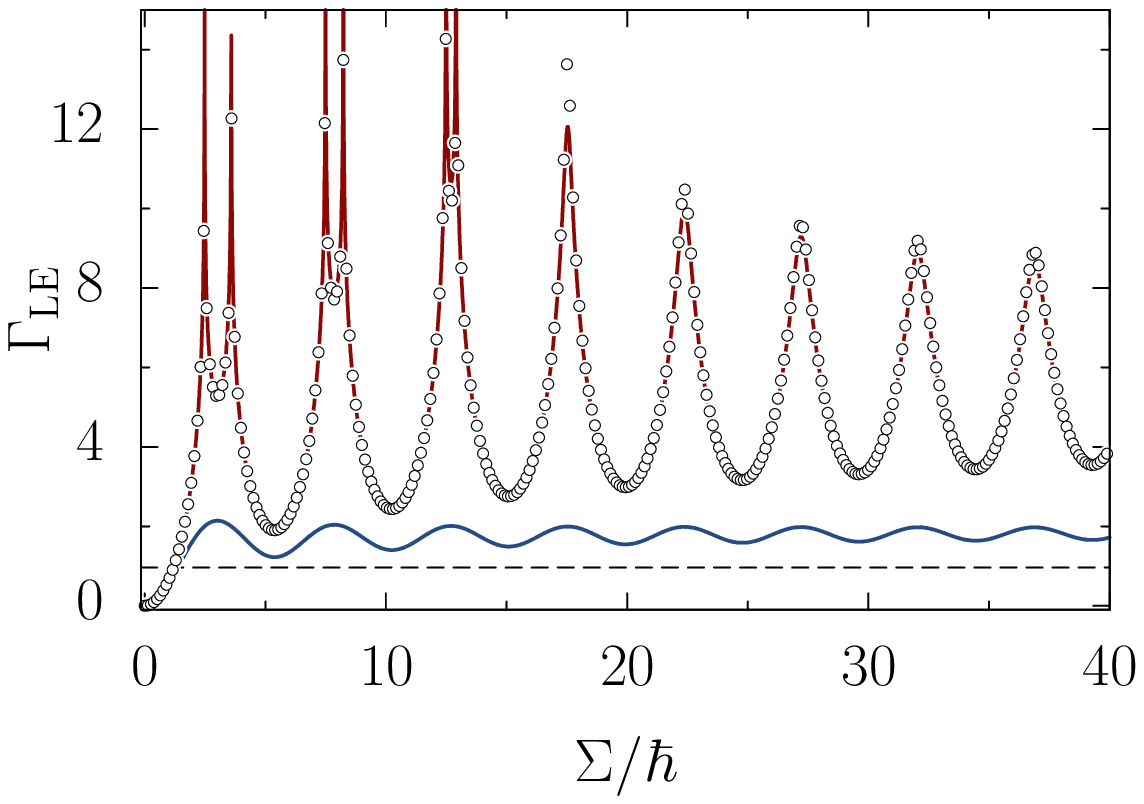} 
\caption{(colour online).
The white circles indicate the 
decay rate $\Gamma_{\rm LE}$ as a function of 
the rescaled perturbation $\Sigma/\hbar$ for 
the quantum perturbed cat map with $a=b=1$, 
$N=2^{12}$. 
Average is done over $\sim 10^3$ initial position states. 
The red line corresponds to $\GAFADR$, 
while the dashed blue line indicates  $\sigma_{_{\rm LDOS}}^{{\rm sc}}$. 
The dashed horizontal (black) line indicates $\lambda\approx 0.96$.
\label{fig:pos}
}
\end{center}
\end{figure}
\subsection{Extended initial states}
\label{sec:ext}
{In the previous section the numerical results show clearly that the
exponential decay of the LE for large $\lambda$ depends crucially on $\GAFADR$. 
In this section we show 
this can also be the case for small $\lambda$. This happens
when the type of initial state chosen has a large linear extension.
Already at the beginnings of LE research, the universality of the perturbation independent regimes 
was questioned and shown to be deeply dependent on the type of initial state
 \cite{Diego2002}. Here we show for classically meaningful initial states
 \cite{Jacquod2009}, like position states, and for squeezed Gaussian states that the decay rate of the LE is given by $\GAFADR$
even for 
small values of $\lambda$.

 We first choose position states. On the quantized torus position states can be pictured as strips of length one and width $\hbar/(2\pi)$.
 In figure~\ref{fig:pos} we computed $\GLE$ 
 }where the initial states 
used were random position states. 
The map is the perturbed cat map described at the beginning of Sec.~\ref{sec:numres}, with $a=b=1$ corresponding to $\lambda\approx 0.96$.
We observe that $\GLE$ obtained coincides almost perfectly with $\GAFADR$. 
For the sake of clarity  and contrast we also 
included the plot for $\sldos$ (solid blue line).
We remark that in the case of extended initial states our conjecture that $\GAFA$ of \equa{eq:gammafull} gives 
the correct decay rate of the LE is fulfilled for initial times, for arbitrary 
values of $\lambda$ and $\alpha$. We would like to also point out that after a transient time, if the Hilbert space is large enough, then a second regime with decay rate given by $\lambda$ might also be observed. 
However for large $\lambda$ the Lyapunov regime becomes increasingly difficult to observe 
because of the saturation at $1/N$.

\begin{figure}[htbp]
\begin{center}
\includegraphics[width=0.8\linewidth]{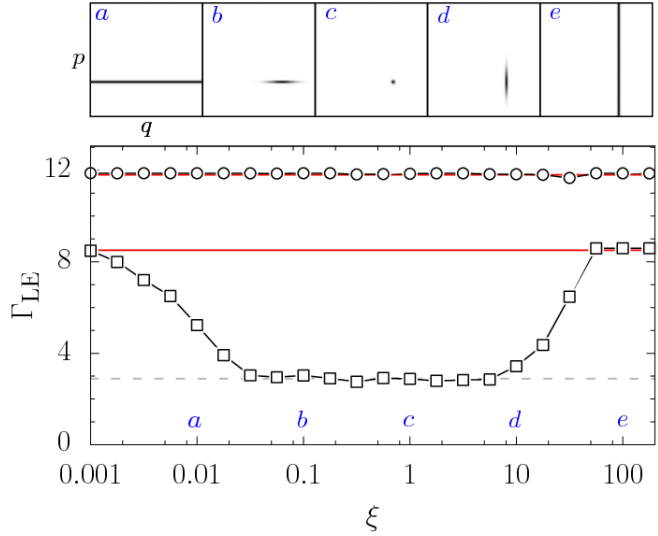} 
\caption{
$\Gamma_{\rm LE}$ as a function of the squeezing parameter $\xi$ for $N=2^{14}$, $a=b=4$ 
and two different values of $\Sigma/\hbar$: 
($\square$) $\Sigma/\hbar=2.467$; ({\Large $\circ$}) $\Sigma/\hbar=46.55$.
The average was done over $n_r=10^3$ inital states.
The dashed line indicates $\lambda=2.887$. The solid horizontal (red) lines indicate the corresponding
value of $\GAFADR$ for each perturbation.
The panels on the top show the Husimi function of the  initial states 
corresponding 
selected values of the squeezing parameter $\xi$: (a) 0.01; (b) 0.1; (c) 1; (d) 10; (e) 100. 
The axes range of both $(q,p)$ in the Husimi function plots is $[0,1]$.
\label{fig:slice}
}
\end{center}
\end{figure}
Another way to illustrate the dependence on the initial state is by varying $\xi$ in \equa{eq:coh}. We then get squeezed
coherent states. If  $\xi=1$ we have the case of Sect.~\ref{sec:Gauss}.
For $\xi<1$ the state squeezes vertically and stretches horizontally, approaching a momentum state for $\xi\ll 1$;  for $\xi>1$ 
the state  squeezes horizontally and stretches vertically, approaching a position state for $\xi\gg 1$. 
In figure~\ref{fig:slice} we show $\Gamma_{\rm LE}$ as a function of the squeezing parameter $\xi$ 
--starting from a 
momentum state  [$\xi\ll 1$], 
passing through a circular coherent state [$\xi=1$], 
over to a position state [$\xi\gg 1$]. We plotted this for 
 $a=b=4$ ($\lambda\approx 2.887$), $N=2^{14}$ and
$10^3$ initial states in the case of perturbation $f_1(q)$ [\equa{eq:pert}] . We took two different values of the rescaled perturbation: $\Sigma/\hbar=2.467$ ($\square$) corresponds in figure~\ref{fig:eco}
to a region where $\GLE$ follows $\GAFADR$ and thus the value of the decay rate remains approximately constant; 
$\Sigma/\hbar=46.55$ ({\Large $\circ$}) corresponds to a region where $\Gamma_{\rm LE}$ has already settled around $\lambda$.  In this case we observe a transition from a decay rate dominated by the AFA for $\xi\lesssim0.1$, $\xi\gtrsim10$, i.e. where the coherent state has been stretched in
either in $q$ or in $p$ by a factor of the order $10$.
\section{Conclusions}
\label{sec:conc}
The Loschmidt echo is a quantity which was introduced as
a measure of the instability and irreversibility of a system under an external perturbation.
The Lyapunov regime of the LE
describes how for some parameter values the decay becomes independent
of the perturbation and the decay rate is the Lyapunov exponent of the classical system. 
It is a well known theoretical prediction {that can be seen in simulations [e.g. for the kicked top \cite{Jacquod2001}, in the stadium billiard \cite{pulpo2002}, or a two-dimensional Lorentz gas\cite{Cucc2004}] }
but { which for some  systems 
that exhibit generic chaotic behaviour -- like quantum maps -- can be}
truly difficult to find
due to very strong deviations \cite{Andersen2006,Silvestrov2003,Wang2004,Pozzo2007,Natalia2009,Casabone2010} .
In this work we have thus set out to try to understand  
this elusive nature of the Lyapunov regime in quantum maps. 
For these systems we show that
 the AFA plays a 
significant role and we dare conjecture is a greater 
defining feature of a system. A similar conclusions could be inferred from previous works on 
billiards with perturbed walls \cite{Goussev2008,Diego2010}. We think the extension of our
results
to more realistic models, which fulfill the uniformly hyperbolic  and chaotic condition, should be straightforward.

We have derived a semiclassical expression $\GAFADR$ for the
decay rate of the AFA in the case of highly chaotic maps (large $\lambda$ limit).
Using extensive numerical simulations in quantum maps
we have shown that,
the decay rate of the LE is very well reproduced by $\GAFADR$ for some range of 
perturbation values.
As a consequence,  
there are three regimes depending on the perturbation. For small $\Sigma/\hbar$ there 
is the well known FGR regime. Then there is an oscillating behaviour which can be explained 
with the decay rate of the AFA and finally for very large $\Sigma/\hbar$ there is the Lyapunov regime.
We point out that the decay rate obtained
for the AFA correctly explains the oscillations that are observed numerically
in this paper but which had been also observed previously 
\cite{Wang2004,Pozzo2007,Natalia2009,Casabone2010}.
The complete understanding of the crossover between the last two regimes is left for further studies \cite{pulpo2011}.

We have also compared $\GAFADR$ to the width of the LDOS and indicated
their fundamental differences when the perturbation is not localized. Moreover,
we have shown that the influence of $\GAFADR$ in the decay of the LE is enhanced when the
initial states are extended in phase space.

We remark that the reasons we find here to the absence of a Lyapunov
regime are fundamentally different to those found in \cite{Silvestrov2003} where
the fluctuations due to different ways of averaging are shown to lead to 
a double-exponential decay of the LE.

\section*{Acknowledgments}
We acknowledge the support from CONICET (PIP-112- 200801-01132) and UBACyT (X237).
I. G.-M. and D. A. W. are researchers of CONICET. Discussions with E. G. Vergini and H. M. Pastawski
are thankfully acknowledged. I.G.-M. thanks partial financial support by the PEPS--PTI from the CNRS (France) and hospitality at the Laboratoire de Physique Th\'eorique (CNRS), Univerit\'e Toulouse III, and from the grant UNAM-PAPIIT IN117310  (Mexico) as well as hospitality at Instituto de F\'isica at UNAM (Mexico).

\section*{References}
\providecommand{\newblock}{}

\end{document}